\def\cal#1{{\cal #1}}

\def\m@th{\mathsurround=0pt}
\def\n@space{\nulldelimiterspace=0pt \m@th}
\def\biggg#1{{\mbox{$\left#1\vbox to 20.5pt{}\right.\n@space$}}}

\def\beginenum{\begin{enumerate}}
\def\endenum{\end{enumerate}}
\def\bitem{\begin{itemize}}
\def\eitem{\end{itemize}}
\def\bray{\begin{array}}
\def\eray{\end{array}}
\def\begindoc{\begin{document}}
\def\enddoc{\end{document}}
\def\bq{\begin{equation}}
\def\eq{\end{equation}}
\def\bqy{\begin{eqnarray}}
\def\eqy{\end{eqnarray}}
\def\bqyn{\begin{eqnarray*}}
\def\eqyn{\end{eqnarray*}}
\def\bc{\begin{center}}
\def\ec{\end{center}}
\def\bfll{\begin{flushleft}}
\def\efll{\end{flushleft}}
\def\bflr{\begin{flushright}}
\def\eflr{\end{flushright}}
\newcommand{\Avec}{\mbox{\boldmath $A$}}
\newcommand{\Bvec}{\mbox{\boldmath $B$}}
\newcommand{\Evec}{\mbox{\boldmath $E$}}
\newcommand{\Fvec}{\mbox{\boldmath $F$}}
\newcommand{\Gvec}{\mbox{\boldmath $G$}}
\newcommand{\Rvec}{\mbox{\boldmath $R$}}
\newcommand{\Uvec}{\mbox{\boldmath $U$}}
\newcommand{\Vvec}{\mbox{\boldmath $V$}}
\newcommand{\evec}{\mbox{\boldmath $e$}}
\newcommand{\jvec}{\mbox{\boldmath $j$}}
\newcommand{\kvec}{\mbox{\boldmath $k$}}
\newcommand{\nvec}{\mbox{\boldmath $n$}}
\newcommand{\uvec}{\mbox{\boldmath $u$}}
\newcommand{\vvec}{\mbox{\boldmath $v$}}
\newcommand{\wvec}{\mbox{\boldmath $w$}}
\newcommand{\xvec}{\mbox{\boldmath $x$}}
\newcommand{\omegavec}{\mbox{\boldmath $\omega$}}
\newcommand{\Omegavec}{\mbox{\boldmath $\Omega$}}

\documentclass[twocolumn,article]{revtex4}%
\usepackage{amsfonts}
\usepackage{amsmath}
\usepackage{amssymb}
\usepackage{color}
\usepackage{graphicx}%
\begin{document}

\title{Nonlinear Coupling of Electromagnetic and Electron Acoustic Waves in
Multi-Species Degenerate Astrophysical Plasma}

\author{N. L. Shatashvili}
\email{nana.shatashvili@tsu.ge} \affiliation{Andronikashvili Institute of
Physics, TSU, Tbilisi 0177, Georgia }
\affiliation{Department of
Physics, Faculty of Exact and Natural Sciences, Ivane
Javakhishvili Tbilisi State University (TSU), Tbilisi 0179,
Georgia}
\author{S. M. Mahajan}
\email{mahajan@mail.utexas.edu} \affiliation{Institute for Fusion
Studies, The University of Texas at Austin, Austin,Tx 78712}
\author{V. I. Berezhiani}
\email{vazhab@yahoo.com} \affiliation{Andronikashvili Institute of
Physics, TSU, Tbilisi 0177, Georgia }
\affiliation{School of
Physics, Free University of Tbilisi, Georgia}

\pacs{47.75.+f, 52.27.Ny, 52.30.Ex, 52.35.Mw, 52.35.We, 97.10.Ex, 97.20.Pm, 97.20.Rp}

\begin{abstract}

Nonlinear wave--coupling is studied in a multi-species
degenerate astrophysical plasma consisting of two electron species
(at different temperatures): a highly degenerate main component
plus a smaller classical relativistic flow immersed
in a static neutralizing ion background.
It is shown that the high frequency electromagnetic (HF EM) waves,
through their strong nonlinear interactions with the electron--acoustic
waves (sustained by a multi-electron component (degenerate) plasma
surrounding a compact astrophysical object) can scatter to lower frequencies so that
the radiation observed faraway will be spectrally shifted downwards.
It is also shown that, under definite conditions, the EM waves
could settle into stationary Solitonic states.  It is expected
that the effects of such structures may persist as detectable signatures in
forms of modulated micro-pulses in the radiation observed far away
from the accreting compact object. Both these effects will
advance our abilities to interpret the radiation coming out of the compact objects.

\end{abstract}

\startpage{1}
\endpage{1}
\maketitle

\section{Introduction}

In compact astrophysical objects (like neutron stars, white dwarfs (WD),
active galactic nuclei (AGN)), the matter density
is so high that the average inter-particle distance is considerably
smaller than the electron De Broglie wave-length.
The electron gas at such high densities must obey Fermi-Dirac
statistics \citep{Compact}. Within the framework of an ideal Fermi gas, several authors have,
recently, studied the nonlinear dynamics of such highly degenerate systems.
One must also note that the Fermi momentum of electrons \ $p_{F} = m_{e}c\,(n_d/n_{cr})^{1/3}$ \
($n_{cr} = 5.9\times 10^{29}\,cm^{-3}$ \ is the normalizing critical
number-density (see e.g. \citep{Akbari,BSM_deg} and references therein) )
acquires highly relativistic values ($\gg m_{e}c$) if \ $n_d\gg n_{cr}$ \ .
For such a dense, degenerate system, the Fermi energy, rather than
the intrinsic thermal energy (which could be very small) will dominate
the system dynamics. In other words the Fermi temperature is much
larger than the thermal temperature,  \ $T \ll T_F \ \ = m_{e}c^{2}(\gamma_F -1)$ \,
where $\gamma_F=\left( 1+\left(\frac{p_F}{m_ec}\right )^{2}\right)^{1/2}$
Even for the least compact of the compact systems, the white dwarfs (WDs),
the densities exceed \ $n_{cr}$ \ necessitating a relativistic treatment for
dynamics for the degenerate gas.

Most reported work has concentrated on low frequency (LF) longitudinal
plasma mode dynamics in quantum/degenerate magneto-plasmas (see e.g.
\citep{Shukla,Haas-q} and references therein). Since
the generation of high density plasma is presumably augmented by
production of intense pulses of X-- and Gamma--rays, several
authors have also investigated modulational interactions of
such high frequencey (HF) waves with variety of plasma modes
\citep{HK,degenerate,BS-self,misra-1,rozina,chanturia,goshadze,mikaberidze}.
It was argued that the observed radiation coming from the
compact astrophysical objects (harboring multi species plasma)
could carry footprints/information from nonlinear interactions
like, for instance, the wave self-modulation and soliton formation.

In the present study, we investigate the nonlinear coupling of the
Electromagnetic (EM) and Electron Acoustic Waves (EAW) in
a Multi-Species Degenerate Astrophysical Plasma consisting of
two different temperature electron species: a highly degenerate
main component ($d$) mixed with a smaller classical relativistic
flow ($cl$) immersed in a static neutralizing ion background.
One of the principal aims of this work  is to explore,
possibly, novel  nonlinear wave-coupling and modulational interactions
induced by the new physics originating in the contamination by the component \ $cl$ .
Such a composite system of a highly degenerate
WD plasma co-existing with a classical hot accreting
astrophysical flow (\citep{mukai,2TDeg} and references therein) is an interesting
and unusual state of matter. It is expected that this
very combination of $d$ and $cl $ will pertain, for example,
during the relativistic jet formation from accretion-induced collapsing
White Dwarfs to Black Holes \citep{Begelman,Kryvdyk,JetsWD}.

Two temperature plasmas have been extensively studied in the past
in the context of  electron-sound wave generation in both classical
and degenerate/quantum plasmas
\citep{Ang,Armstrong,Eka,Barnes,Feldman-1,Feldman-2,Feldman-3,misra,MM,masood,Sah,2Tquantum}.
Normally the cooler component is a smaller fraction with lower
density compared to the hotter component \citep{Eka,Barnes}.
Nonlinear phenomena in multi-component plasmas can be seriously
affected by relativistic temperatures. An example of how a two
temperature e–-p–-i plasma can differ from a single temperature
system was worked out in  \citep{2T-epi} where it was shown that
the presence of a minority of cold electrons and ions can lead
to the scattering of the pump EM wave into the electron-sound
and EM waves, and  to the instability of relativistically hot
e–-p plasma against the LF perturbations. EAW could also exist
in two temperature plasmas consisting of a degenerate
(fermi temperature $T_F$) and a classical component ($T_{cl}\ll T_F$).
The dispersion properties of EAW can be derived from
the linear dispersion relation shown, e.g., in \citep{Haas},
modified by the presence of non-degenerate cold electron
species (static ions):
\begin{equation}
1 + \frac{3\Omega_{ed}^2}{k^2V_{Fe}^2} \left[1 - \frac{1}{2}
\frac{\omega}{kV_{Fe}} ln\left( \frac{\omega + kV_{Fe}}{\omega - kV_{Fe}} \right) \right]
+ \frac{\omega_{ecl}^2}{\omega^2} = 0 \ ,
\label{dispersion}
\end{equation}
where \ $V_{Fe}=p_{F}/m_e\,\gamma_F$ \ is the relevant relativistic Fermi velocity,
$\Omega_{ed}=\omega_{ed}/\gamma_F$ \, ( $\omega_{ed}=\sqrt{{4\pi e^2N_{0d}/m_e}}$)
\ and  $\omega_{ecl}=\sqrt{{4\pi e^2N_{0cl}/m_e}}$ are, respectively, the plasma
frequencies of the two electron species), $\gamma_F =\sqrt{1+R_0^2}$, $R_0 =
\left( {N_{0d}/n_{cr}} \right)^{1/3} $, and $N_{0d}$ ($N_{0cl}$) is
equilibrium lab-frame density of the degenerate (classical) plasma.
We have assumed $T_{ecl} \equiv 0$.

When Landau-damping for degenerate electrons is neglected,
Equation (\ref{dispersion}), in the limit,
\begin{equation}
kV_{ecl} \ll \omega \ll kV_F,
\label{Sound-limits}
\end{equation}
reduces to the so-called Electron-Sound solution with the frequency
\begin{equation}
\omega^2 = \omega_{ecl}^2\,\frac{k^2V_F^2}{3\Omega_{ed}^2}
\left[1 + \frac{k^2V_F^2}{3\Omega_{ed}^2}  \right]^{-1} =
k^2 c_s^2 \left[1+\frac{k^2c_s^2}{\Omega_{ed}^2} \right]^{-1} \ ,
\label{SoundF}
\end{equation}
where
\begin{equation}
c_s^2 = \frac{\omega_{ecl}^2}{3\Omega_{ed}^2}\,V_F^2 = \frac{c^2}{3}\,
\left(\frac{N_{0cl}}{N_{0d}}\right) \,\frac{R_0^2}{\sqrt{1+R_0^2}}\ ,
\label{SoundV}
\end{equation}
which, due to the condition (\ref{Sound-limits}), is satisfied for
\begin{equation}
\frac{N_{0cl}}{N_{0d}} \ll \frac{3}{\sqrt{1+R_0^2} } \ .
\label{condition}
\end{equation}
We note that in above dispersion relation, the quantum effects 
(like recoil and pair creation)
are neglected \citep{Melrose,Haas,2Tquantum}.

\section{Model}

The nonlinear wave dynamics will be studied for a quasi neutral
unmagnetized plasma of an immobile classical ion component ($i$),
and two electron species -- the bulk relativistic degenerate ($d$)
electron gas with a density $N_{0d}$ and a small contamination
of non-degenerate classical ($cl$) electrons with
density $N_{0cl}$ \citep{2TDeg}. Quasi neutrality demands
\begin{equation}
N_{0d} + N_{0cl} = N_{0i} \ \ \Longrightarrow \ \
\frac{N_{0i}}{N_{0d}} = 1+\alpha, \quad   \alpha \equiv \frac{N_{0cl}}{N_{0d}} \  ,
\label{B-eq}
\end{equation}
where \ $\alpha \,(\ll 1)$ \ is the fraction of the classical to the degenerate electrons.

\bigskip

To study the nonlinear propagation of intense EM waves, the multicomponent
plasma dynamics has to be coupled (see \citep{2TDeg}) with the
Maxwell equations,
\begin{equation}
{\bf E} = -\frac{1}{c}\,\frac{\partial }{\partial t}{\bf A} - \nabla \varphi \ ,
\qquad {\bf B} = \nabla \times {\bf A} \ ,
\label{Potentials}
\end{equation}
take the form (in the Coulomb Gauge $\nabla \cdot {\bf A} = 0$):
\begin{equation}
\frac{\partial ^{2}\mathbf{A}}{\partial t^{2}} - c^{2}\Delta \mathbf{A} +
c\,\frac{\partial }{\partial t}\left( \mathbf{\nabla }\varphi \right) -4\pi c\,
\mathbf{J}=0 \ ,
\label{S1}
\end{equation}
\begin{equation}
\Delta \varphi =-4\pi \,\rho \ ,
\label{S2}
\end{equation}
where for charge density (sum over all species) and current density
(sum over only the two electron species) we have, respectively:
\[
\rho =\sum q\,N = - eN_{d} - eN_{cl} + eN_{i} \ ,
\]
\begin{equation}
\mathbf{J}=\sum q\gamma_{d(cl)} n_{d(cl)}\mathbf{V }_{d(cl)}= - \frac{e}{m} N_{d}
\frac{\mathbf{p}_{d}}{\gamma _{d}}-\frac{e}{m}N_{cl}\frac{\mathbf{p}_{cl}}{\gamma _{cl}} .
\label{S3}
\end{equation}
Here \ ${\bf p}_{d(cl)}= \gamma_{d(cl)} m{\bf
V}_{d(cl)}$ \ is the hydrodynamic momentum, \ $n_{d(cl)} =
N_{d(cl)}/\gamma_{d(cl)}$ \ is the rest-frame particle density \
($N_{d(cl)}$ \  denotes the laboratory frame density) \
of the degenerate (classical) electron fluid element, \
${\bf V}_{d(cl)}$ \ is the fluid velocity, and \
$\gamma_{d(cl)} =\left( 1+{\bf p}_{d(cl)}^{2}/m^2c^{2}\right)^{1/2}$.

The fully covariant description of the active fluid species (ions are static)
is displayed in its familiar vortical form, for example, in \citep{M-min, M-EV, BM-94}.
Since we will be dealing with a particular class of systems for which the
generalized (canonical) vorticities \  ${\bf \Omega}_{d(cl)} = - (e/c )
{\bf B+\nabla\times}\left( G_{d(cl)}{\bf p}_{d(cl)}\right) = 0$, the remanent
dynamics is contained in (see \citep{BM-94} for details)
\begin{equation}
\frac{\partial }{\partial t}\left( G_{d(cl)}\mathbf{p}_{d(cl)}-\frac{e}{c}\mathbf{A}%
\right) +\mathbf{\nabla }\left( mc^{2}\,G_{d(cl)}\,\gamma _{d(cl)}-e\,\varphi
\right) =0\ ,
\label{S4}
\end{equation}
\begin{equation}
\frac{\partial N_{d(cl)}}{\partial t} \ + \ \nabla \cdot ({N_{d(cl)}{\bf
V}_{d(cl)}})=0 \ , \label{Cont}
\end{equation}
where the "effective mass factors" $G_{d}$ and $G_{cl }$ are quite
different for the two electron species:
\ $G_{d} = w_{d}/n_{d}m_{e}c^{2}$ , \ where \ $w_{d}$ \ is
an enthalpy per unit volume, originates from degeneracy
rather than relativistic kinematics. The general expression for enthalpy
\ $w_{d}$ \ for arbitrary density and temperature (for a plasma
described by local Dirac-Juttner equilibrium distribution
function) can be found in \citep{boltzmann}. For a fully (strongly)
degenerate electron plasma, however, this very
tedious expression smoothly transfers to the one with just density
dependence: $w_{d}\equiv w_{d}(n)$ \citep{BSM_deg}. In
fact \ $w_{d}/n_{d}m_{e}c^{2}=\left(
1+(R_{d})^{2}\right)^{1/2}$, where \ $R_d$ [$=
(n_d/n_{cr})^{1/3}$]. The effective mass factor,
then, is simply determined by the plasma rest frame density,
$G_{d}=[ 1+(n_{d}/n_{cr})^{2/3}]^{1/2}$ \ for arbitrary \ $n_{d}/n_{cr}$ .
For relativistically hot classical plasma an expression for effective mass
factor $G_{cl}$ can be found in \citep{BM-94,Ryu}. Note, that
the degenerate fluid equation (\ref{S4}) is valid in the long wave-length limit ($k\ll 1$);
the characteristic frequencies are much greater than the De Broglie frequency:
$\omega \gg \omega_{\hbar}={\hbar}\,k^2/2m$ justifying the absence of
quantum diffraction phenomenon effects \citep{Rukhadze}.

\bigskip

We will, now,  investigate the one-dimensional propagation of
circularly polarized EM wave ($\partial_z \neq 0, \ \partial_x \equiv 0,
\ \partial y \equiv 0$) with a mean frequency $\omega_0$ and a
mean wave number $k_0$ in the $z$ direction:
\begin{equation}
{\bf A}_{\perp} = (\hat{\bf x} + i\ {\hat{\bf y}})\,A(z,t)\,exp(ik_0z -i\omega_0t) + c.c.
\label{S6}
\end{equation}
where $A(z,t)$ is a slowly varying function of $z$ and $t$ and
$\hat{\bf x}$ and $\hat{\bf y}$ are the standard unit vectors.
The vanishing of the generalized vorticities ( ${\bf \Omega}_{d(cl)}=0$ )
guarantees
\begin{equation}
\mathbf{p}_{\perp \,d(cl)}=\frac{e}{c\,G_{d(cl)}}\,\mathbf{A}_{\perp } \ ,
\label{S7}
\end{equation}
where the constant of integration is set equal to zero since particle hydrodynamic
momenta are assumed to be zero at the infinity where the field vanishes.

\bigskip

The perpendicular component of Eq.(\ref{S1}), using Eq.-s (\ref{S3} - \ref{S7}),
will give the Equation for transverse motion as follows:
\[
\frac{\partial^{2}\mathbf{A}_{\perp }}{\partial t^{2}}-c^{2}\frac{\partial
^{2}\mathbf{A}_{\perp }}{\partial z^{2}}
\]
\begin{equation}
+\omega _{ed}^{2}\left( \frac{N_{d}}
{N_{d0}}\frac{1}{\gamma _{d}G_{d}}\mathbf{+}\frac{N_{cl}}{N_{d0}}\frac{1}
{\gamma_{cl}G_{cl}}\right) \mathbf{A}_{\perp }  =0 \ ,
\label{S8}
\end{equation}
where
\begin{equation}
\omega_{ed} = \sqrt{\frac{4\pi e^2 N_{0d}}{m}} \ , \qquad 
\gamma _{d(cl)}=\sqrt{1+\frac{e^{2}\mathbf{A}_{\perp }^{2}}{m^{2}c^{4}G_{d(cl)}^{2}}} \ ,
\label{S9}
\end{equation}
\[
G_{d}=\left[1+R_{0}^{2}\left(\frac{N_{d}}{\gamma_{d}N_{d0}}\right)^{2/3}
\right]^{1/2} \ , 
\]
\begin{equation}
R_{0}=\left(\frac{N_{d0}}{n_{cr}}\right)^{1/3} \ ,
\qquad G_{cl} = G_{cl}\left( mc^{2}/T_{cl}\right) \ .
\label{S10}
\end{equation}

\bigskip

We also need to write the equations for longitudinal motion. This motion
is driven by the ponderomotive pressure ($ \sim {{\bf p}_{\perp \,d(cl)}}^2 $ )
of high frequency (HF) EM wave. Equations (\ref{Cont}) and (\ref{S4}), then, give :
\begin{equation}
\frac{\partial }{\partial t}N_{d(cl)}+\frac{\partial }{\partial z}N_{d(cl)}
\frac{p_{zd(cl)}}{m_{e}\gamma _{d(cl)}}=0 \ ,
\label{S11}
\end{equation}
\begin{equation}
\frac{\partial }{\partial t}\left( G_{d(cl)}p_{zd(cl)}\right) +\frac{\partial }
{\partial z}\left( mc^{2}\,G_{d(cl)}\,\gamma _{d(cl)}-e\,\varphi \right) =0 \
\label{S12}
\end{equation}
and Eq.(\ref{S2}) reads as:
\begin{equation}
\Delta \varphi = 4\pi e\left( N_{d} + N_{cl}-N_{i} \right) \ .
\label{S13}
\end{equation}

In what follows we assume that the pump wave is weakly relativistic
($|{\bf A}|^2 \ll m^2c^2$) leading to the expansion of
densities so that $\delta N_{d(cl)} \ll N_{0d(cl)} $ .
Straightforward algebra gives the following relation
for the degenerate fluid:
\[
\frac{1}{G_{d}\,\gamma _{d}}=\frac{1}{\sqrt{R_{0}^{2}+1}}
-\frac{R_{0}^{2}}{3\left( R_{0}^{2}+1\right) ^{3/2}}\frac{\delta N_{d}}{N_{0d}}
\]
\begin{equation}
- \frac{1}{2}\frac{1}{\left( R_{0}^{2}+1\right)^{3/2}}\frac{e^{2}\mathbf{A}%
_{\perp }^{2}}{m_{e}^{2}c^{4}}\left( 1-\frac{1}{3}\frac{R_{0}^{2}}{%
R_{0}^{2}+1}\right) \
\label{S14}
\end{equation}
while for the classical non-degenerate electron fraction we obtain:
\begin{equation}
\frac{1}{G_{cl}\,\gamma_{cl}}=\frac{1}{G_{0cl}}\left( 1-\frac{1}{2}\frac{e^{2}
\mathbf{A}_{\perp }^{2}}{m^{2}c^{4}G_{0cl}^{2}}B_{1}+\frac{\delta N_{cl}}
{N_{0cl}}B_{2}\right) \ ,
\label{S15}
\end{equation}
where
\begin{equation}
B_{1}=\left( 1+\frac{1}{2}\frac{G_{cl}^{\prime }\left( z_{0}\right) }{%
f^{\prime }(z_{0})G_{0cl}^{3}}\right) \ , \qquad
B_{2}=\frac{G_{0cl}^{\prime }f\left( z_{0}\right) }{G_{0cl}^{2}f^{\prime }(z_{0})} \ .
\label{S16}
\end{equation}
In the preceding equations, the prime denotes the $z$-derivative, and
the effective mass of the  classical electrons species is ($K_2(z)$ and $K_3(z)$
are modified Bessel functions)
\begin{equation}
G_{cl} = \frac{K_{3}( z_{cl}) }{K_{2}( z_{cl}) }
\qquad  {\rm{with}} \qquad  z_{cl} = mc^{2}/T_{cl} \ ,
\label{S17}
\end{equation}
and
\[
\frac{N_{cl}}{\gamma_{cl}}f( z_{cl}) = const \ , \quad
\frac{N_{cl}}{\gamma_{cl}}f(z_{cl}) = N_{0cl}f(z_{0}) \ , 
\]
\begin{equation}
f(z_{cl}) = \frac{z_{cl}}{K_{2}(z_{cl})}\exp -z_{cl}G_{cl}(z_{cl}) \ .
\label{S18}
\end{equation}

Using the notations
\begin{equation}
\frac{\delta N_{d}}{N_{d0}} = \delta \nu_{d} \ , \qquad
\frac{\delta N_{cl}}{N_{0cl}} = \delta \nu_{cl}  \ ,
\label{S19}
\end{equation}
we may express the set of equations for HF and LF motions in a ``simplified" form:
\[
\frac{\partial ^{2}\mathbf{A}_{\perp }}{\partial t^{2}}-c^{2}\frac{\partial
^{2}\mathbf{A}_{\perp }}{\partial z^{2}}
\]
\begin{equation}
+ \omega _{ed}^{2}\left[ \left(
1+\delta \nu _{d}\right) \frac{1}{\gamma _{d}G_{d}}\mathbf{+}\alpha \left(
1+\delta \nu_{cl}\right) \frac{1}{\gamma _{cl}G_{cl}}\right] \mathbf{A}_{\perp} = 0 \ ,
\label{S20}
\end{equation}

\[
\left( 1+\delta \nu _{d}\right)\ \frac{\omega _{ed}^{2}}{\gamma _{d}G_{d}}
=\Omega _{d}^{2} 
\]
\begin{equation}
+ \Omega _{d}^{2}\left[ \delta \nu _{d} - \frac{1}{2}
\frac{1}{\left( R_{0}^{2}+1\right) }\frac{e^{2}\mathbf{A}_{\perp }^{2}}
{m^{2}c^{4}}\right] \left( 1 - \frac{1}{3}\frac{R_{0}^{2}}{R_{0}^{2} + 1}\right) \ ,
\label{S21}
\end{equation}

\[
\alpha \, \left( 1+\delta \nu _{cl}\right) \frac{\omega _{ed}^{2}}{\gamma
_{h}G_{cl}}
\]
\begin{equation}
=\alpha \,\Omega_{cl}^{2}+\alpha \,\Omega _{cl}^{2}\left[\delta
\nu_{cl}\left( B_{2}+1\right) -\frac{1}{2}\frac{e^{2}\mathbf{A}_{\perp }^{2}}{%
m^{2}c^{4}G_{0cl}^{2}}B_{1}\right] \ ,
\label{S22}
\end{equation}
where
\begin{equation}
\Omega _{d}^{2} = \frac{\omega _{ed}^{2}}{\sqrt{R_{0}^{2}+1}} \ , \qquad
\qquad \Omega_{cl}^{2} = \frac{\omega_{ed}^{2}}{G_{0cl}}  \ .
\label{S23}
\end{equation}
Equation for the HF motion, after some simple algebra, can be simplified to obtain
\[
\frac{\partial ^{2}\mathbf{A}_{\perp }}{\partial t^{2}} -
c^{2}\frac{\partial^{2}\mathbf{A}_{\perp }}{\partial z^{2}}
+ \left( \Omega_{d}^{2}+\alpha \,\Omega_{cl}^{2}\right) \mathbf{A}_{\perp }
\]
\[
+\Omega_{d}^{2} \left( 1-\frac{1}{3}\frac{R_{0}^{2}}{R_{0}^{2}+1}\right)
\left[ \delta \nu_{d}-\frac{1}{2}\frac{1}{\left(
R_{0}^{2}+1\right) }\frac{e^{2}\mathbf{A}_{\perp }^{2}}{m^{2}c^{4}}
\right]\mathbf{A}
_{\perp }
\]
\begin{equation}
+ \alpha \,\Omega _{cl}^{2}\left[ \delta \nu_{cl}\left( B_{2}+1\right) -\frac{1}
{2}\frac{e^{2}\mathbf{A}_{\perp }^{2}}{m^{2}c^{4}G_{0cl}^{2}}B_{1}\right]
\mathbf{A}_{\perp } = 0  \ .
\label{S24}
\end{equation}

\bigskip

The plasma response is contained in Eqs. (\ref{S2} - \ref{S4}):
\begin{equation}
\frac{\partial ^{2}}{\partial z^{2}}\delta \varphi =4\pi eN_{0d}\left(
\delta \nu_{d}+\alpha \delta \nu_{cl}\right) \ ,
\label{S25}
\end{equation}
\begin{equation}
\frac{\partial }{\partial t}\delta \nu _{d(cl)}+\frac{\partial }{\partial z}
\frac{\delta p_{zd(cl)}}{m}=0 \ .
\label{S26}
\end{equation}
After some tedious but straightforward algebra we derive the following
equation for, respectively, the degenerate and non-degenerate classical fluids, :
\[
\frac{\partial }{\partial t}\frac{\delta p_{zd}}{m}-\frac{e}{m
\sqrt{R_{0}^{2}+1}}\frac{\partial }{\partial z}\ \delta \varphi
+ \frac{c^{2}}{3}\frac{R_{0}^{2}}{R_{0}^{2}+1}\frac{\partial }
{\partial z}\ \delta \nu_{d}
\]
\begin{equation}
+ \ \frac{1}{2}\frac{c^{2}}{R_{0}^{2} + 1}
\left( 1-\frac{1}{3}\frac{R_{0}^{2}}{\left( R_{0}^{2}
+ 1\right) }\right) \frac{\partial }{\partial z}\left( \frac{e^{2}
\mathbf{A}_{\perp }^{2}}{m^{2}c^{4}}\right) = 0 \ ,
\label{S27}
\end{equation}

\[
\frac{\partial }{\partial t}\frac{\delta p_{zcl}}{m}-\frac{e}{mG_{0cl}}%
\frac{\partial }{\partial z}\ \delta \varphi -c^{2}B_{2}\frac{\partial }
{\partial z}\ \delta \nu_{cl}
\]
\begin{equation}
+ \frac{c^{2}}{2}\frac{B_{1}}{G_{0cl}^{2}}
\frac{\partial }{\partial z}\left( \frac{e^{2}\mathbf{A}_{\perp }^{2}}
{m^{2}c^{4}}\right) =0\ .
\label{S28}
\end{equation}

Introducing $B_{2}=-\left\vert B_{2}\right\vert$, using (\ref{S26}),
Eqs. (\ref{S27}) and (\ref{S28}) can be rewritten in a somewhat
more elegant way,
\begin{equation}
\frac{\partial ^{2}}{\partial t^{2}}\ \delta \nu_{d} \ +
\label{S29}
\end{equation}
\[
\frac{\partial }{\partial z}\left[ \frac{e}{m\sqrt{R_{0}^{2}+1}}
\frac{\partial }{\partial z}\ \delta \varphi -\frac{c^{2}}{3}\frac{R_{0}^{2}}{R_{0}^{2}+1}
\frac{\partial }{\partial z}\ \delta \nu_{d} \right]
\]
\[
- \frac{1}{2}\,\frac{\partial }{\partial z}\left[ \frac{c^{2}}{R_{0}^{2}+1}
\left( 1-\frac{1}{3}\frac{R_{0}^{2}}{\left( R_{0}^{2}+1\right) }\right)
\frac{\partial }{\partial z}\left( \frac{e^{2}\mathbf{A}_{\perp }^{2}}
{m^{2}c^{4}}\right) \right] =0
\]

\noindent and
\[
\frac{\partial }{\partial t}\frac{\delta p_{zcl}}{m}-\frac{e}{mG_{0cl}}
\frac{\partial }{\partial z}\,\delta \varphi + c^{2}\left\vert
B_{2}\right\vert \frac{\partial }{\partial z}\ \delta \nu_{cl}
\]
\begin{equation}
+\frac{c^{2}}{2}
\frac{B_{1}}{G_{0cl}^{2}}\frac{\partial }{\partial z}\left( \frac{e^{2}
\mathbf{A}_{\perp }^{2}}{m^{2}c^{4}}\right) =0 \ .
\label{S30}
\end{equation}

\bigskip

Introducing
\begin{equation}
\kappa ^{2}=\left( 1-\frac{1}{3}\frac{R_{0}^{2}}{\left( R_{0}^{2}+1\right)} \right)
\ , \qquad \frac{2}{3} <  \kappa^2 < 1
\label{S31}
\end{equation}
we finally write the set of equations for coupled LF motion:
\[
\left( \frac{\partial ^{2}}{\partial t^{2}}-\frac{c^{2}}{3}\frac{R_{0}^{2}}{%
R_{0}^{2}+1}\frac{\partial ^{2}}{\partial z^{2}}+\frac{\omega _{ed}^{2}}{%
\sqrt{R_{0}^{2}+1}}\right) \ \delta \nu_{d}
\]
\begin{equation}
-\frac{1}{2}\frac{\kappa ^{2}c^{2}}
{R_{0}^{2}+1}\frac{\partial ^{2}}{\partial z^{2}}\left( \frac{e^{2}
\mathbf{A}_{\perp }^{2}}{m^{2}c^{4}}\right) 
=-\frac{\alpha \, \omega_{ed}^{2}}
{\sqrt{R_{0}^{2}+1}}\ \delta \nu_{cl} \ ,
\label{S32}
\end{equation}

\[
\left( \frac{\partial^{2}}{\partial t^{2}}-c^{2}\left\vert B_{2}\right\vert
\frac{\partial^{2}}{\partial z^{2}}+\frac{\alpha \, \omega_{ed}^{2}}{G_{0cl}}
\right) \ \delta \nu_{cl}-\frac{c^{2}}{2}\frac{B_{1}}{G_{0cl}^{2}}
\frac{\partial^{2}}{\partial z^{2}}\left( \frac{e^{2}\mathbf{A}_{\perp }^{2}}{%
m^{2}c^{4}}\right) 
\]
\begin{equation}
= - \frac{\omega_{ed}^{2}}{G_{0cl}}\ \delta \nu_{d} \ .
\label{S33}
\end{equation}

\bigskip

\noindent From the definition of classical hot electrons effective
mass $G_{cl}$ we may show that generally $B_2 < 0$.

\bigskip

Our system of equations actually describes the plasma with two different
effective masses of electron species -- effective mass of degenerate electrons
($G_{0d}(N_{0d})$) is determined by their density while the classical
hot electrons' effective mass is determined by their temperature ($G_{0cl}(T_{0cl})$).
Although, the derived system is valid for arbitrary temperatures of the
classical component, we will limit ourselves (consistent with our earlier
stated goal) to the case when the effective mass of degenerate
electrons is considerably larger than the classical fraction -- the Fermi
energy of degenerate electrons is significantly bigger than the thermal energy
of the classical electron fraction.  For the $d$ component, then,
the factors $B_1\simeq 1$, and $B_2\simeq 0$ . Consequently, after invoking,
(\ref{condition}), the equations describing the LF motion, may be written as:
\[
\left[ \Omega _{d}^{2} \frac{\partial
^{2}}{\partial t^{2}}-\alpha \,\omega _{ed}^{2}\frac{c^{2}}{3}\frac{R_{0}^{2}}{%
R_{0}^{2}+1}\frac{\partial ^{2}}{\partial z^{2}}\right] \ \delta \nu
_{d}=
\]
\begin{equation}
= \ - \ \alpha \ \Omega _{d}^{2}\frac{c^{2}}{2}\left( 1-\frac{\kappa ^{2}}{\sqrt{%
R_{0}^{2}+1}}\right) \frac{\partial ^{2}}{\partial z^{2}}\left( \frac{e^{2}%
\mathbf{A}_{\perp }^{2}}{m^{2}c^{4}}\right)
\label{S35}
\end{equation}
and
\begin{equation}
\delta \nu_{cl}=-\frac{1}{\alpha }\delta \nu_{d}+\frac{c^{2}}{2\alpha
\omega _{ed}^{2}}\frac{\partial ^{2}}{\partial z^{2}}\left( \frac{e^{2}
\mathbf{A}_{\perp }^{2}}{m^{2}c^{4}}\right) \ .
\label{S37}
\end{equation}
In the low frequency regime:
\begin{equation}
\frac{\partial ^{2}}{\partial t^{2}}\ll \alpha \, \omega _{ed}^{2} \ll \Omega_{ed}^2 \
\qquad
\label{S34}
\end{equation}
calculating the combination (from the preceding)
\[
\Omega_{d}^{2}\,\kappa^{2}\,\delta \nu_{d} + \alpha \,\omega_{ed}^{2}\,\delta \nu_{cl}
\]
\begin{equation}
= - \omega_{ed}^{2}\left( 1-\frac{\kappa^{2}}{\sqrt{R_{0}^{2}+1}}\right)
\delta \nu_{d}+\frac{c^{2}}{2}\frac{\partial^{2}}{\partial z^{2}}\left(
\frac{e^{2}\mathbf{A}_{\perp }^{2}}{m_{e}^{2}c^{4}}\right),
\label{S38}
\end{equation}
we obtain the HF equation for the vector potential in its final form:
\[
\frac{\partial ^{2}\mathbf{A}_{\perp }}{\partial t^{2}}-c^{2}\frac{\partial
^{2}\mathbf{A}_{\perp }}{\partial z^{2}}+\left( \Omega_{d}^{2} + \alpha \,
\omega_{ed}^{2}\right) \mathbf{A}_{\perp } 
\]
\[
- \omega_{ed}^{2}\left( 1
- \frac{\kappa^{2}}{\sqrt{R_{0}^{2}+1}}\right) \delta \nu _{d}\mathbf{A}_{\perp }
+ \frac{c^{2}}{2}\mathbf{A}_{\perp }\frac{\partial^{2}}{\partial z^{2}}\left(
\frac{e^{2}\mathbf{A}_{\perp }^{2}}{m_{e}^{2}c^{4}}\right) -
\]

\begin{equation}
-\frac{\Omega_{d}^{2}\,\kappa ^{2}}{2\left( R_{0}^{2}+1\right) }\frac{e^{2}
\mathbf{A}_{\perp }^{2}}{m_{e}^{2}c^{4}}\mathbf{A}_{\perp }
- \frac{\alpha \, \omega_{ed}^{2}}{2}\frac{e^{2}\mathbf{A}_{\perp }^{2}}{m^{2}c^{4}}
\mathbf{A}_{\perp }=0 \ .
\label{S39}
\end{equation}
In above equation the non--local (5th) term is much smaller
than the diffraction one and it can be ignored.

\bigskip

By imposing the  HF  dispersion relation: \ $\omega_0^2 = k_0^2c^2 +
\Omega_d^2 + \alpha \omega_{ed}^2$ \ , the final coupled
HF/LF dynamics is expressible as (in terms of the dimensionless $A \equiv eA/mc^2 $
and for arbitrary level of degeneracy):
\[
2i \omega_0 \left( \frac{\partial }{\partial t}
+ V_g\frac{\partial}{\partial z} \right) A
+\omega_0 V_g^{\prime }\frac{\partial^{2} A} {\partial z^{2}}
\]
\[
+ \omega_{ed}^{2} \left( 1-\frac{\kappa^{2}}{\sqrt{R_{0}^{2}+1}}\right)
\delta \nu_{d}\,A -
\]

\begin{equation}
+ \omega_{ed}^2\left( \alpha +\frac{\kappa ^{2}}{\left(
R_{0}^{2}+1\right) ^{3/2}}\right) |A|^2 A = 0 \ ,
\label{S40}
\end{equation}

\[
\left(  \frac{\partial^{2}}{\partial t^{2}}-c_s^{2}\frac{\partial^{2}}
{\partial z^{2}}\right) \delta \nu_{d}
\]
\begin{equation}
= - 3\ c_s^2 \ \frac{\sqrt{R_{0}^{2}+1}}{R_0^2}\,
\left( 1-\frac{\kappa^{2}}{\sqrt{R_{0}^{2}+1}}\right)
\frac{\partial^2|A|^2}{\partial z^2} \ ,
\label{S41}
\end{equation}
where $V_g$ is the group velocity of HF wave and
the electron sound velocity $c_s$ is defined as
\begin{equation}
c_s^2 = \alpha \,\frac{c^2}{3}\,
\frac{R_0^2}{\sqrt{R_0^2 + 1}} \ .
\label{S42}
\end{equation}

From (\ref{S42}), we can read the relativistic electron-sound
for a 2-temperature degenerate relativistic plasma,
velocity $c_s \sim \sqrt{\frac{\alpha}{3}}\,c\,R_0 /
(R_0^2 + 1)^{1/4} < \ c/\sqrt{3} $ \ .

We emphasize here, that the presence of the non-degenerate
electron fraction is crucial, without a non zero $\alpha $ ,
the speed would not be finite. In fact,  \ $\alpha \gg c_s^2
/ L^2\omega_{0cl}^2$ \ where $L$ is the characteristic length
of LF mode. It must be emphasized that the presence of a small
fraction of non-degenerate electrons is the  reason that
LF longitudinal waves exist together with the HF--EM waves.

\section{Nonlinear Coupling of HF EM waves and Electron-Acoustic Waves}

The system of Equations (\ref{S40}), (\ref{S41}) together with
the relation (\ref{S38}), constitutes the  system that we now
investigate for a possible modulational instability arising from
the interaction of HF--EM pump wave with the LF the electron-sound waves.

We begin by rewriting the relevant equations as
\[
2i \omega_0 \left( \frac{\partial }{\partial t}
+ V_g\frac{\partial}{\partial z} \right) A
+\omega_0 V_g^{\prime }\frac{\partial^{2} A} {\partial z^{2}}
\]
\begin{equation}
+ \omega_{ed}^{2} \ b_1 \ \delta \nu_{d}\,A + \omega_{ed}^2 \ b_2
|A|^2 \,A = 0 \ ,
\label{S43}
\end{equation}

\begin{equation}
\left(  \frac{\partial^{2}}{\partial t^{2}}-c_s^{2}\frac{\partial^{2}}
{\partial z^{2}}\right) \delta \nu _{d}
= - 3\ c_s^2 \ b_3 \frac{\partial^2|A|^2}{\partial z^2}
\label{S44}
\end{equation}
with
\[
b_1 = \left[ 1-\frac{\kappa^{2}}{\sqrt{R_{0}^{2}+1}}\right] \ ,
\quad  b_2 = \left[ \alpha +\frac{\kappa ^{2}}{\left(R_{0}^{2}+1\right) ^{3/2}}\right]  \ ,
\]
\begin{equation}
b_3 = \frac{\sqrt{R_{0}^{2}+1}}{R_0^2}\ b_1
\label{S45}
\end{equation}

\bigskip

and making the ansatz
\[
A(z,t) = a(z,t) = \ e^{i\theta(z,t)} \ ; 
\]
\[
\delta \nu_d(z,t) =
\delta \nu_d(z,t)\ exp[ikz - i\Omega t] \ + \ c.c. \ ;
\]
\[
a = a_0 + \delta a \ exp[ikz - i\Omega t] \ + \ c.c. \ ; 
\]
\begin{equation}
\theta = \theta_0 + \delta \theta \ exp[ikz - i\Omega t] \ + \ c.c. \ ,
\label{S46}
\end{equation}
where $a(z,t)$ and $\theta(z,t)$ are slowly varying functions
of space and time, and $\delta a\ll a_0 \ , \ \delta \theta \ll \theta_0$.
From the resulting linearized Eqs.  (\ref{S43} - \ref{S45}),
we obtain the dispersion relation:
\[
\left( \Omega ^{2}-c_{s}^{2}k^{2}\right) \left[ \left( \Omega
-V_{g}k\right) ^{2}-\frac{1}{4}V_{g}^{^{\prime }}k^{2}\left( V_{g}^{^{\prime
}}k^{2}-2b_{2}\frac{\omega _{ed}^{2}}{\omega _{0}}a_{0}^{2}\right) \right]
\]
\begin{equation}
= \frac{3}{2}\,b_{1}\,b_{3}\,V_{g}^{^{\prime }}\,c_{s}^{2}\,k^{4}a_{0}^{2}
\ \frac{\omega _{ed}^{2}}{\omega _{0}} \ .
\label{S47}
\end{equation}

For coinciding roots and  for small amplitudes
(with $\Omega^{2} - c_{s}^{2}k^{2} \ll 1 $), we get:
\[
\left( \Omega ^{2}-c_{s}^{2}k^{2}\right) \left[ \left( \Omega
-V_{g}k\right) ^{2}-\frac{1}{4}V_{g}^{^{\prime }2}k^{4}\right]
\]
\begin{equation}
= \frac{3}{2}b_{1}b_{3}V_{g}^{^{\prime }}c_{s}^{2}k^{4}a_{0}^{2}
\frac{\omega _{ed}^{2}}{\omega _{0}} \ ,
\label{S55}
\end{equation}
which with
\[
\Omega = kc_{s}+i\Gamma \ , \qquad \qquad
\Omega = V_{g}k-\frac{1}{2}V_{g}^{\prime }k^{2}+i\Gamma
\]
leads to
\begin{equation}
\left( \Omega^{2}-c_{s}^{2}k^{2}\right) \left[ \left( \Omega -V_{g}k\right)
^{2}-\frac{1}{4}V_{g}^{^{\prime }2}k^{4}\right] = 2kc_{s}V_{g}^{\prime
}k^{2}\Gamma^{2}
\label{S56}
\end{equation}
implying the increment of decay instability:
\begin{equation}
\Gamma^{2} = \frac{3}{4}\,c_{s}\,k\,b_{1}\,b_{3}\,a_{0}^{2}\,
\frac{\omega_{ed}^{2}}{\omega _{0}} \ .
\label{S57}
\end{equation}

The principal message of the preceding calculation is that
in the multi-species (degenerate in bulk) plasma it is possible
to generate lower frequency electromagnetic waves when high transverse
HF EM waves scatter on the longitudinal electron-sound;
the corresponding relative frequency shift \ $\Delta \,\omega /\omega_0 \sim
c_s/c = \sqrt{\alpha/3}\,R_0/(1+R_0^2)^{1/4}$ \ being defined
by the classical electron fraction \ $\alpha = N_{0cl}/N_{0d}$ \ as well as the
degeneracy level $R_0(N_{0d})$ of bulk electron species. E.g. for intermediate
degeneracy level $R_0 \sim 1$ we get for such shift to be $\sim 0.5 \sqrt{\alpha }$.

\bigskip

Now let's look for the modulational instability assuming
$\Omega \simeq V_{g}k + i\Gamma,  \ V_{g}k \gg \Gamma $ , for
the growth rate of instability we obtain:
\[
\Gamma ^{2} = \frac{1}{4}V_{g}^{^{\prime }}k^{2}\left[ 2\,\frac{\omega
_{ed}^{2}}{\omega _{0}}\,a_{0}^{2}\,B - V_{g}^{^{\prime }}k^{2}\right] \ \ ,
\]
\begin{equation}
\qquad \rm{with} \qquad
B \equiv  b_{2}-3\,b_{1}\,b_{3}\,\frac{c_{s}^{2}}{V_{g}^{2}-c_{s}^{2} } \ \ .
\label{S48}
\end{equation}
For the modulational instability \ $B>0$ ; \ this translates into
\begin{equation}
c_{s}^{2}\left( 1+3\frac{b_{1}b_{3}}{b_{2}}\right) < V_{g}^{2} < c_{s}^{2} \ \ .
\label{S49}
\end{equation}
The maximum growth rate occurs at
\[
k_{m}^{2} = \frac{\omega_{ed}^{2}}{V_{g}^{^{\prime }}\omega_{0}}\,a_{0}^{2}\,B \
\]
and is
\begin{equation}
\Gamma_{m}^{2}=\frac{1}{4}\,V_{g}^{^{\prime }}\,k_{m}^{2}\,\left[ 2\,\frac{\omega
_{de}^{2}}{\omega_{0}}\,a_{0}^{2}\,B-V_{g}^{^{\prime }}\,k_{m}^{2}\right] \
= \frac{1}{2}\,V_{g}^{^{\prime }}\,k_{m}^{2} \ ,
\label{S50}
\end{equation}
with the explicit expression
\begin{equation}
\Gamma_{m}=\frac{\omega_{ed}}{2}\left( \frac{\omega _{ed}}{\omega _{0}}
\right) a_{0}^{2}\left( b_{2}-3b_{1}b_{3}\frac{c_{s}^{2}}{V_{g}^{2}-c_{s}^{2}}\right) \ .
\label{S51}
\end{equation}
We must remind the reader that since our final equations were derived under
the condition (\ref{S34}),  the wave-amplitude can not be very large:
\begin{equation}
\Gamma_{m} \ll \sqrt{\alpha }\,\omega_{ed} \qquad \Longrightarrow \qquad
a_{0}^{2} \ll \sqrt{\alpha} \,\frac{\omega_{0}}{\omega_{ed}B} \ .
\label{S52}
\end{equation}

\bigskip

Let us make some estimates:

\noindent (i) For super relativistic degenerate bulk electrons ($R_0 \gg 1$), we find
\[
b_{1}=\left( 1-\frac{2}{3R_{0}}\right) =1 \ , \quad b_{2}
= \left( \alpha +\frac{2}{3R_{0}^{3}}\right) \ ,
\]
\[
b_{3}=\frac{1}{R_{0}} \qquad \Longrightarrow
\qquad B \cong \alpha + \frac{3}{R_{0}} \
\]
and the amplitude is restricted to:
\begin{equation}
a_{0}^{2} < \sqrt{\alpha }\, \frac{\omega_{0}}{\omega_{ed}}
\left( \alpha + \frac{3}{R_{0}}\right)^{-1} \ .
\label{S53}
\end{equation}

\noindent (ii) For weakly relativistic degeneracy ($R_0 \ll 1$),
the relevant results are
\[
b_{1}=1 \ , \qquad b_{2} = ( 1 + \alpha ) \ , \qquad b_{3} =
\frac{1}{R_{0}^{2}} \gg 1 \qquad
\Longrightarrow 
\]
\[
\qquad B \cong \frac{3}{R_{0}^{2}} \gg 1
\]
and the amplitude is bounded by
\begin{equation}
a_{0}^{2} \ll \sqrt{\alpha }\,R_{0}^{2}\,\frac{\omega_{0}}{\omega_{ed}} \ .
\label{S54}
\end{equation}

\bigskip

Thus, the addition of even very small amount of non-degenerate
classical plasma ($N_{0cl}\neq 0$, i.e. $c_s \neq 0$) leads
to the instability of degenerate e--i plasma against the LF perturbations.
Note, that the wave modulation/decay instabilities,
demonstrated above, do not exist in an e–-i plasma that
has a single electron (degenerate) component.

\bigskip

For a better understanding of the character of radiation
coming from compact objects, let us explore possible stationary
solutions to the system of Equations (\ref{S40}), (\ref{S41}). Let
\[
{\bf A}_{\perp}={\bf A}_{\perp}(\xi, \tau) \ , \quad \delta \nu_d
= \delta \nu_d (\xi, \tau) \ ,
\]
\begin{equation}
\xi = z - V_g t \ , \quad t = \tau , \ \quad
\frac{\partial }{\partial \tau} \ll V_g\frac{\partial }{\partial z} \ .
\label{S58}
\end{equation}
Solving (\ref{S44}) yields:
\begin{equation}
\delta \nu_d = - 3 b_3 \,\frac{c_s^2}{V_g^2 - c_s^2}\,|A|^2 \ ,
\label{S59}
\end{equation}
and using it in (\ref{S43}) gives the so called Nonlinear Schr\"{o}dinger (NLS) equation:

\begin{equation}
2i \omega_0 \frac{\partial }{\partial \tau} A
+\omega_0 V_g^{\prime }\frac{\partial^{2} A} {\partial \xi^{2}}
+ \omega_{ed}^{2} \ Q\,|A|^2 A = 0 \ ,
\end{equation}
\[
Q =  \frac{3c_s^2}{c_s^2 - V_g^2}\,b_1 b_3 + b_2 \ .
\label{S60}
\]

As it is well known the NLS allows a stationary solution at $Q > 0$ representing:
1) the subsonic ($V_g < c_s$) soliton of rarification (the total density variation
$\delta \nu_d + \delta \nu_h \sim \delta \nu_h < 0$ since $\alpha \ll 1$ -- see
Eq. (\ref{S37}) ). For the supersonic regime ($ \sqrt{1+\frac{3b_1b_3}{b_2}}\,c_s < V_g \ll V_F$)
we get the soliton of compression.



\section{Conclusions}

High density (Fermi degenerate) plasmas associated with compact
objects are generated simultaneously with the  production of
intense pulses of X-- and Gamma--rays. Recently
it was argued that these strong HF electromagnetic waves
could play a fundamental role in shaping/controlling the
nature of the final radiation that will emerge from these
compact astrophysical objects. Amongst several possibilities,
we study here a subclass of phenomena induced by the interaction
of the HF waves with the plasma; the  relevant plasma
consists of two electron components (in a neutral background
of static ions) -- the bulk component is  high density (degenerate
with relativistic Fermi energies/relativistic Fermi temperature $T_F$)
and is ``contaminated'' with a much lower density classical
(non degenerate) component.

All the new and significant results of this paper stem for our
inclusion of the classical component that is relatively
low density, and has much smaller  temperature $T$ (thermal)
as compared to $T_F$. Because of the strong nonlinear coupling
between the  HF EM waves and and Electron Acoustic Waves
(a normal mode of the plasma), this ``multi-electron"
system is found to display a modulation/decay instability,
and the HF EM waves can be strongly scattered
on the acoustic waves. The main contribution of this
work is the demonstration that it is the small classical
contamination that is the source of this instability;
such an instability will not pertain in a system that
has a single active electron component.

Since lower frequency electromagnetic waves can be generated
by the scattering of HF EM waves on the electron--acoustic waves, one
expects that the electromagnetic spectral range can become
quite different from the original; in fact the radiation coming out
of the compact object will be driven  spectrally downwards.

We also found that the system does allow stationary
soliton solution both in the subsonic and supersonic regimes,
respectively,  the solitons of rarification and compression.
One could expect that the effects of such solitonic structures
may persist as detectable signatures in forms of modulated micro-pulses
\citep{kennel} in the radiation far away from the accreting compact object.

It is well known that most of the compact
astrophysical objects are immersed in strong magnetic fields.
It is highly desirable, therefore, to extend the
present study to magnetized multi-component plasmas.
The importance of the spin and the quantum diffraction effects
must be carefully estimated, and if necessary, must be included
in the model. These modifications/improvements of the model
will be essential before we undertake  parametric / numerical
studies so that the model could be compared to real observations
on compact astrophysical objects.

\section{Acknowledgements}

Authors acknowledge the support from Shota Rustaveli Georgian
National Foundation Grant Project No. FR17-391.
Work of SMM was supported by US DOE Contract No.DE-FG02-04ER54742.


\begin{thebibliography}{99}                                                                                               %


\bibitem{Compact} L. Shapiro \& S. A. Teukolsky. Black
Holes, White Dwarfs and Neutron Stars: The Physics of Compact
Objects (John Wiley and Sons, New York, 1973).

\bibitem{Akbari} M. Akbari-Moghanjoughi. Phys. Plasmas \textbf{20},
042706 (2013).

\bibitem{BSM_deg} V.I. Berezhiani,  N.L. Shatashvili \& S.M. Mahajan.
Phys. Plasmas {\bf 22}, 022902 (2015).

\bibitem{Shukla} P.K. Shukla \& B. Eliasson. Rev. Mod. Phys. {\bf
83}, 885 (2011).

\bibitem{Haas-q} F. Haas. Quantum Plasmas: An Hydrodynamic Approach. Springer
Science \& Business Media, {\bf 65} (2011).

\bibitem{HK} F. Haas \& I. Kourakis. Plasma. Phys. Control. Fusion {\bf 57}, 044006
(2015).

\bibitem{degenerate} V.I. Berezhiani, N.L. Shatashvili  \&
N.l. Tsintsadze. Physica Scripta {\bf 90(6)}, 068005 (2015).

\bibitem{BS-self} V.I. Berezhiani \& N.L. Shatashvili. Phys. Plasmas
{\bf 23}, 104502 (2016).

\bibitem{misra-1} A. P. Misra \& Debjani Chatterjee. Phys.
Plasmas {\bf 25}, 062116 (2018).

\bibitem{rozina} Ch. Rozina,  S. Ali, N. Maryam \& N. L. Tsintsadze. Phys.
Plasmas {\bf 25}, 093302 (2018).

\bibitem{chanturia} G. T. Chanturia,  V. I. Berezhiani \& S. M. Mahajan,  Phys.
Plasmas {\bf 24}, 074501 (2017.)

\bibitem{goshadze} M. Goshadze, V.I. Berezhiani, Z. Osmanov Phys. Lett. A
{\bf 383(10,11)}, 1027 (2019).

\bibitem{mikaberidze} G. Mikaberidze \& V.I. Berezhiani. Physics Letters A
{\bf 379(42)}, 2730 (2015).

\bibitem{mukai} K. Mukai.  {\it PASP} {\bf 129}, 062001 (2017)

\bibitem{2TDeg} N.L. Shatashvili,  S.M. Mahajan,  V.I. Berezhiani. AAS,
{\bf 364}, 148 (2019).

\bibitem{Begelman} M.C. Begelman,  R.D. Blandford \&
M.D. Rees. \ Rev. Mod. Phys. {\bf 56} 255 (1984).

\bibitem{Kryvdyk} V. Kryvdyk. {\bf MNRAS}, {\bf 309}, 593 (1999).

\bibitem{JetsWD}  V. Kryvdyk \& A. Agapitov.
15th European Workshop on White Dwarfs
ASP Conference Series, {\bf 372}, 411 (2007).

\bibitem{Ang} L.K. Ang, P. Zhang. Phys. Rev. Lett.
{\bf 98}, 164802 (2007).

\bibitem{Armstrong} R.J. Armstrong, W.J. Weber,
J. Trulsen. Phys. Lett. {\bf 74A}, 319–322
(1979).

\bibitem{Barnes} W.L. Barnes, A. Dereux, T.W. Ebbesen. Nature (London)
{\bf 424}, 824 (2003).

\bibitem{Eka} E.M. Khirseli \& N.L. Tsintsadze. Fizika Plazmy {\bf 6}, 1081; 1980,
Sov. J. Plasma Phys. {\bf 6}, 595 (1990).

\bibitem{Feldman-1} W.C. Feldman, J.R. Asbridge, M.D. Montgomery, S.P. Gary. J. Geophys.
Res. {\bf 80}, 4181 (1975).

\bibitem{Feldman-2} W.C. Feldman, R.C. Anderson, S.J. Bame, S.P. Gary, J.T. Gosling,
D.J. McComas, M.F. Thomsen, G. Paschmann, M.M. Hoppe.
J. Geophys. Res. {\bf 88}, 96 (1983).

\bibitem{Feldman-3} W.C. Feldman, R.C. Anderson, S.J. Bame, J.T. Gosling, R.D. Zwickl.
 J. Geophys. Res. {\bf 88}, 9949 (1983).

\bibitem{misra}  A. P. Misra, P. K. Shukla \& C. Bhowmik.
Phys. Plasmas {\bf 14}, 082309 (2007).

\bibitem{MM} S. Mahmood \& W. Masood. Phys. Plasmas {\bf 15}, 122302 (2008).

\bibitem{masood} W. Masood \& A. Mushtaq. Phys. Plasmas {\bf 15}, 022306 (2008).

\bibitem{Sah} O. P. Sah \& J. Manta. Phys. Plasmas {\bf 16}, 032304 (2009).

\bibitem{2Tquantum} S. Chandra \& B. Ghosh. AAS, {\bf 342},
417 (2012).

\bibitem{2T-epi} N.L. Shatashvili, J.I. Javakhishvili \&
H. Kaya. AAS, {\bf 250}, 109 (1997).

\bibitem{Haas} F. J. Haas. Plasma Phys. {\bf 82}, 705820602 (2016).

\bibitem{Melrose} D. Melrose. Quantum Plasmadynamics: Unmagnetized Plasmas.
Springer (2008).

\bibitem{M-min} S.M. Mahajan, \prl {\bf 90}, 035001 (2003).

\bibitem{M-EV} S.M. Mahajan. Phys. Plasmas, {\bf 23(11)},
112104 (2016).

\bibitem{BM-94} V.I. Berezhiani \& S.M. Mahajan.
Phys. Rev. Lett. {\bf 73}, 1110 (1994); \ Phys. Rev. E {\bf 52}
1968 (1995).

\bibitem{boltzmann} C. Cercignani \& G. M. Kremer. \textit{The relativistic
Boltzmann equation: theory and applications}, Birkh\"{a}user,
Basel, (2002).

\bibitem{Ryu} D. Ryu, I. Chattopadhyay \& E. Choi.
J. Korean Phys. Soc. {\bf 49(4)}, 1842 (2006).

\bibitem{Rukhadze} M. V. Kuzelev, A.A. Rukhadze. Physics Uspekhi, {\bf 42},
603 (1999).

\bibitem{kennel} A.C.L. Chian \& C.F. Kennel,
Astrophys. Space Sci. {\bf 97}, 9 (1983).


\end{thebibliography}
\end{document}